\documentclass{ifacconf}

\usepackage{graphicx}      
\usepackage{natbib}        
\usepackage{amssymb}
\usepackage{amsmath}
\usepackage{enumerate}
\usepackage{color}
\usepackage{algorithm}
\usepackage{algpseudocode}
\begin{document}
	\begin{frontmatter}
		
		\title{Data-Driven Robust Stabilization with Robust DOA Enlargement for Nonlinear Systems} 

		\author[First]{Chaolun Lu},
		\author[First]{Yongqiang Li},
		\author[Second]{Zhongsheng Hou},
		\author[First]{Yuanjing Feng},
		\author[First]{Yu Feng},
		\author[Third]{Ronghu Chi},
		\author[Fourth]{Xuhui Bu}

		\address[First]{College of Information Engineering, Zhejiang University of Technology, Hangzhou, China; Tel:+086-15657191030(Yongqiang Li), e-mail:luke@zjut.edu.cn(Chaolun Lu), yqli@zjut.edu.cn(Yongqiang Li), fyjing@zjut.edu.cn(Yuanjing Feng), yfeng@zjut.edu.cn(Yu Feng)}
		\address[Second]{School of Automation, Qingdao University, Qingdao, China, (e-mail:zshou@qdu.edu.cn)}
		\address[Third]{School of Automation and Electronic Engineering, Qingdao University of Science and Technology, Qingdao, China, (e-mail:ronghu\_chi@qust.edu.cn)}
		\address[Fourth]{School of Electrical Engineering and Automation, Henan Polytechnic University, Jiaozuo, China, (e-mail: buxuhui@gmail.com)}
		
		\begin{abstract}                
			Most of nonlinear robust control methods just consider the affine nonlinear nominal model. When the nominal model is assumed to be affine nonlinear, available information about existing non-affine nonlinearities is ignored. For non-affine nonlinear system, Li et al. (2019) proposes a new nonlinear control method to solve the robust stabilization problem with estimation of the robust closed-loop DOA (Domain of attraction). However, Li et al. (2019) assumes that the Lyapunov function is given and does not consider the problem of finding a good Lyapunov function to enlarge the estimate of the robust closed-loop DOA. The motivation of this paper is to enlarge the estimate of the closed-loop DOA by selecting an appropriate Lyapunov function. To achieve this goal, a solvable optimization problem is formulated to select an appropriate Lyapunov function from a parameterized positive-definite function set. The effectiveness of proposed method is verified by numerical results. 
		\end{abstract}
		
		
	\end{frontmatter}
	
	\section{Introduction}
	Research on robust control began in the 1960s (\cite{Petersen:2014_1315}). In the past two decades, robust control has been a research hot-spot due to its ability of dealing with the uncertainty. Most of robust control theory is linear (assume that the nominal model is linear) (\cite{Safonov:2012_173,Petersen:2014_1315,Bhattacharyya:2017_45}). Linear robust control ignores available information about existing nonlinearities. When the nonlinearities are significant, the resulting controllers is very conservative. Hence, nonlinear robust control is proposed (\cite{Freeman:2008}).
	
	Most of nonlinear robust control methods just consider the affine nonlinear nominal model that is affine with respect to the control input, such as the Lyapunov min-max approach (\cite{Corless:1993}), the nonlinear H$_\infty$ approach (\cite{Basar:1995}), the robust backstepping approach (\cite{Freeman:2008}) and the approach based on polynomial fuzzy systems (\cite{Ashar:2018_1423,Tsai:2018_3630}). When the nominal model is assumed to be affine nonlinear, available information about existing non-affine nonlinearities is ignored. Considering the discrete-time general (non-affine) nonlinear nominal model, \cite{Li:2019} proposes a new nonlinear robust control method based on Lyapunov function. Due to the difficulty to achieve the global stabilization for general nonlinear systems, this method also estimates the robust closed-loop DOA besides designing a robust controller. However, \cite{Li:2019} assumes that the Lyapunov function is given and does not consider the problem of finding a good Lyapunov function to enlarge the estimate of the closed-loop DOA.
	
	Inspired by \cite{Li:2019}, we observe that, for different Lyapunov functions, the estimates of the closed-loop DOA are totally different. Therefore, the motivation of this paper is to enlarge the estimate of the closed-loop DOA by selecting an appropriate Lyapunov function. In order to achieve this goal, a solvable optimization problem is formulated to selecting an appropriate Lyapunov function from a parameterized positive-definite function set, which is a subset of all sum-of-square polynomials. The objective function of this optimization problem is the volume of the estimate of the closed-loop DOA, which can be obtained based on the method proposed in \cite{Li:2019}. The analytical expression of the volume of the estimate of the closed-loop DOA is hard to be derived, but it is easy to evaluate its value for a given positive-definite function. Hence, meta-heuristic optimization methods can be used to solve this optimization problem, such as particle swarm optimization.
	
	The rest of this paper is organized as follows. In Section 2, the control problem is formulated. In Section 3, the method proposed in \cite{Li:2019} is briefly introduced, which solves the robust stabilization problem with estimation of the closed-loop DOA when the Lyapunov function is given. In Section 4, a solvable optimization problem is formulated and solved to select an appropriate Lyapunov function from a parameterized positive-definite function set. In Section 5, simulation results show the effectivenss of the proposed method. Finally, the conclusion is drawn in Section 6.
	
	Notation: For $x\in\mathbb{R}^n$ and $u\in\mathbb{R}^m,(x;u)$ represents a new vector in $\mathbb{R}^{n+m}$. For $x_1,x_2\in\mathbb{R}^n,x_1<x_2$ means $x_1$ is less than or equal to $x_2$ element by element.
	
	\section{Problem formulation}
	
	Consider the nonlinear discrete-time plant set
	\begin{eqnarray}
	& \!\! \mathfrak{F} := \Big\{ f: \mathbb{R}^n \!\times\! \mathbb{R}^m \!\to \!	\mathbb{R}^n \Big| f(0,0) = 0, \nonumber \\ 
	& \quad \hat{f}(x,u) - \delta(x,u) \leq f(x,u) \leq \hat{f}(x,u) + \delta(x,u) \Big\}, \label{eq:plant_set}
	\end{eqnarray}
	where $x(k) \in \mathbb{R}^n$ is the state, $u(k) \in \mathbb{R}^m$ is the control input, $\hat{f}: \mathbb{R}^n \times \mathbb{R}^m \to \mathbb{R}^n$ is the known nominal model satisfying $\hat{f}(0,0) = 0$ and $\delta: \mathbb{R}^n \times \mathbb{R}^m \to \mathbb{R}^n_+$ is the known modeling error bound satisfying $\delta(0,0) = 0$.The control objective is to find a robust controller $\mu: \mathbb{R}^n \to \mathbb{R}^m$ and an estimate of the robust closed-loop DOA to ensure that, $\forall f \in \mathfrak{F}$, the closed-loop $x(k+1) = f(x(k),\mu(x(k)))$ is asymptotically stable at the origin for all initial states in the estimate of the robust closed-loop DOA and to enlarge the estimate of the robust closed-loop DOA.
	
	The first part of the control objectives, \textit{i.e.}, finding a robust controller and an estimate of the robust closed-loop DOA, can be achieved by using the method proposed in (\cite{Li:2019}). In this paper, the second part of the objective, \textit{i.e.}, enlarging the estimate of the robust closed-loop DOA, is achieved by selecting an appropriate Lyapunov function from a positive-definite function set, \textit{e.g.}, sum-of-square polynomials (\cite{Packard:2010}).

	\section{PRELIMINARIES: robust stabilization with estimation of closed-loop DOA}
	
	In this section, we briefly introduce the method solving robust stabilization with estimation of the closed-loop DOA. See \cite{Li:2019} for more details.
	
	\subsection{Sufficient condition for robust stabilization}
	
	The following theorem gives sufficient conditions for robust stabilization of all plants in the plant set and estimation of the robust closed-loop DOA.
	
	\begin{thm} \label{thm:robust_stab}
		For the plant set $\mathfrak{F}$ defined in \eqref{eq:plant_set}, if a positive-definite function $L: \mathbb{R}^n \to \mathbb{R}$ and a constant $\alpha \in \mathbb{R}_+$ exist such that the level-set $\mathbb{X}_{\mathrm{ls}} (L,\alpha)$ satisfying
		\begin{eqnarray}
		\mathbb{X}_{\mathrm{ls}} (L,\alpha) = \Big\{x \in \mathbb{R}^n \ \Big| \ L(x) \leq \alpha \Big\} \subset \mathbb{X}_{\mathfrak{F}} (L) \cup \{0\}, \label{eq:thm:robust_stab:X_L_alpha}
		\end{eqnarray}
		then any state feedback controller $\mu: \mathbb{R}^n \to \mathbb{R}^m$ satisfying
		\begin{eqnarray}
		\mu(0) = 0, (x;\mu(x)) \in \mathbb{W}_{\mathfrak{F}} (L), \forall x \in \mathbb{X}_{\mathfrak{F}} (L),  \label{eq:thm:robust_stab:mu} 
		\end{eqnarray}
		can asymptotically stabilize any plant $f \in \mathfrak{F}$ for all initial state $x_0 \in \mathbb{X}_{\mathrm{ls}} (L,\alpha)$, where 
		\begin{eqnarray}
		&\mathbb{W}_{\mathfrak{F}} (L) = \Big\{\!(x;\!u) \!\in\! \mathbb{R}^{n\!+\!m} \Big| \forall f \!\in\! \mathfrak{F}, L\big(\!f(x,\!u)\!\big) \!-\! L(\!x\!) \!<\! 0 \Big\}, \label{eq:thm:robust_stab:Omega} \\
		&\mathbb{X}_{\mathfrak{F}} (L) = \Big\{x \in \mathbb{R}^{n} \Big| \exists u \in \mathbb{R}^m, (x;u) \in \mathbb{W}_{\mathfrak{F}}(L) \Big\}. \label{eq:thm:robust_stab:X} 
		\end{eqnarray} 
	\end{thm}
	
	The proof of Theorem~\ref{thm:robust_stab} is presented in (\cite{Li:2019}).
	
	From Theorem~\ref{thm:robust_stab}, we know that any state feedback controller in the robust negative-definite domain(NDD) $\mathbb{W}_{\mathfrak{F}}(L) \subset \mathbb{R}^n \times \mathbb{R}^m$, \textit{i.e.}, satisfying  \eqref{eq:thm:robust_stab:mu}, can asymptotically stabilize all plant in the plant set $\mathfrak{F}$ and that any Lyapunov function level-set $\mathbb{X}_{\mathrm{ls}} (L,\alpha)$ belonging to the robust NDD $\mathbb{X}_{\mathfrak{F}}(L) \subset \mathbb{R}^n$, \textit{i.e.}, satisfying \eqref{eq:thm:robust_stab:X_L_alpha}, can be an estimate of DOA for all closed-loops of all plant in the plant set $\mathfrak{F}$. Hence, if the robust NDDs $\mathbb{W}_{\mathfrak{F}}(L)$ and $\mathbb{X}_{\mathfrak{F}}(L)$ are obtained, it is easy to find a robust controller and an estimate of the robust closed-loop DOA. However, due to nonlinearities of $\hat{f}$,  $\delta$ and $L$, it is hard to obtain analytic solutions of the robust NDDs. In the next subsection, a data-driven method of estimating the robust NDDs is introduced.

	\subsection{Robust negative-definite domains estimation}
	
	Let $\mathbb{X} \subset \mathbb{R}^n$ and $\mathbb{U} \subset \mathbb{R}^m$ denote the interested regions, containing the origins, in the state space and the control space, respectively. Let $\mathbb{W} = \mathbb{X} \times \mathbb{U} \subset \mathbb{R}^{n+m}$ denote the interested region in the state-control space. Then a sample data set $W^d$ of $\mathbb{W}$ can be generated by \eqref{eq:W^d}, in which each data point $(x^d;u^d) \in \mathbb{R}^{n+m}$ is drawn from the uniform distribution on $\mathbb{W}$. 
	\begin{eqnarray}
	& \!\! W^d = \Big\{(x^d_v;u^d_v) \in \mathbb{R}^{n+m}, v = 1,2,\cdots, N^d_{xu} \nonumber \\
	& \qquad \Big| (x^d_v;u^d_v) \sim \mathcal{U} (\mathbb{W}) \Big\}, \label{eq:W^d}
	\end{eqnarray}
	where $N^d_{\text{xu}}$ is the number of data points and $\mathcal{U} (\mathbb{D})$ denotes the uniform distribution on compact domain $\mathbb{D}$ belonging to any multi-dimensional Euclidean space.
	
	Next, we aim to find the sample data set $W^d_{\mathfrak{F}} (L) \subset W^d$ of $\mathbb{W}_{\mathfrak{F}} (L)$ defined in \eqref{eq:thm:robust_stab:Omega}. $\mathbb{W}_{\mathfrak{F}}(L)$ is rewritten as 
	\begin{eqnarray}
	&\!\! \mathbb{W}_{\mathfrak{F}} (L) = \Big\{(x;u) \in \mathbb{R}^{n+m} \Big| \forall \bar{x} \in \bar{\mathbb{X}}_\mathfrak{F} (x,u), \nonumber \\
	& \qquad L(\bar{x}) - L(x) < 0 \Big\}, \label{eq:W_F(L):redefinition}
	\end{eqnarray}
	where $\bar{x} = f(x,u)$ and, for any $(x;u) \in \mathbb{R}^{n+m}$, $\bar{\mathbb{X}}_{\mathfrak{F}} (x,u)$ is defined as
	\begin{eqnarray}
	& \!\!\bar{\mathbb{X}}_\mathfrak{F} (x,u) = \Big\{\bar{x} \in \mathbb{R}^n \Big|  \hat{f}(x,u) - \delta(x,u) \leq \bar{x} \nonumber \\
	& \qquad \leq \hat{f}(x,u) + \delta(x,u) \Big\}. \label{eq:bar{X}_F}
	\end{eqnarray}
	For each data point $(x^d;u^d) \in W^d$ , a sample data set $\bar{X}_{\mathfrak{F}}^d(x^d,u^d)$ of $\bar{\mathbb{X}}_\mathfrak{F} (x^d,u^d)$ is generated by \eqref{eq:bar{X}_F}, in which each data point $\bar{x}^d \in \mathbb{R}^n$ is drawn from the uniform distribution on $\bar{\mathbb{X}}_\mathfrak{F} (x^d,u^d)$.
	\begin{eqnarray}
	& \!\! \bar{X}_{\mathfrak{F}}^d(x^d,u^d) = \Big\{\bar{x}_h \in \mathbb{R}^n, h = 1,2,\cdots, N^d_{\bar{x}} \nonumber \\
	& \qquad \Big| \bar{x}_h \sim \bar{\mathbb{X}}_\mathfrak{F} (x^d,u^d) \Big\}, \label{eq:x_f^d}
	\end{eqnarray}
	where $N^d_{\bar{x}}$ is the number of data points. Hence, from \eqref{eq:W_F(L):redefinition}, the sample data set $W^d_{\mathfrak{F}} (L)$ of $\mathbb{W}_{\mathfrak{F}} (L)$ can be expressed as
	\begin{eqnarray}
	&\!\! W^d_{\mathfrak{F}} (L) = \Big\{(x^d;u^d) \in W^d \Big| \forall \bar{x}^d \in \bar{X}_{\mathfrak{F}}^d(x^d,u^d), \nonumber \\
	& \qquad L(\bar{x}^d) - L(x^d) < 0 \Big\}. \label{eq:W^d_F(L)}
	\end{eqnarray}
	
	Finally, the region $\mathbb{X} \in \mathbb{R}^{n}$ and $\mathbb{U} \in \mathbb{R}^{m}$ are partitioned into disjoint cells. Here, we apply uniform grids over $\mathbb{X}$ and $\mathbb{U}$. Let $\{\mathbb{C}_i^x\}$ denote the partition of $\mathbb{X}$ and $\{\mathbb{C}_j^u\}$ denote the partition of $\mathbb{U}$, where $\mathbb{C}_i^x \subset \mathbb{R}^n, i = 1,2,\cdots, N^C_{\text{x}}$, $\mathbb{C}_j^u \subset \mathbb{R}^m, j = 1,2,\cdots, N^C_{\text{u}}$, and $N^C_{\text{x}}$ and $N^C_{\text{u}}$ are numbers of cells in $\{\mathbb{C}_i^x\}$ and $\{\mathbb{C}_j^u\}$, respectively. Hence, the partition of $\mathbb{W} \in \mathbb{R}^{n+m}$ can be represented as $\{\mathbb{C}^w_t\} = \{\mathbb{C}^x_i\} \times \{\mathbb{C}^u_j\}$, where $\mathbb{C}^w_t \subset R^{n + m}, t = 1,2,\cdots,N^C_{\text{x}} \cdot N^C_{\text{u}}$. With the sample data set $W^d_{\mathfrak{F}} (L)$ of $\mathbb{W}_{\mathfrak{F}}(L)$ and the partition $\{\mathbb{C}^w_t\}$ of $\mathbb{W}$, an estimate $\hat{\mathbb{W}}_{\mathfrak{F}} (L)$ of $\mathbb{W}_{\mathfrak{F}}(L)$ can be obtained by combining all cells in which all data points belong to $W^d_{\mathfrak{F}} (L)$ and is expressed as
	\begin{eqnarray}
	&\hat{\mathbb{W}}_{\mathfrak{F}} (L) \!=\! \Big\{\mathbb{C}^w \in \{\mathbb{C}_t^w\} \Big| \forall (x^d;u^d) \in \mathbb{C}^w \cap W^d, \nonumber \\
	&(x^d;u^d) \in W^d_{\mathfrak{F}} (L) \Big\}. \label{eq:hat{W}_F(L)}
	\end{eqnarray}
	The estimate $\hat{\mathbb{X}}_{\mathfrak{F}}(L)$ of the robust NDD $\mathbb{X}_{\mathfrak{F}}(L)$ (defined in \eqref{eq:thm:robust_stab:X}) can be obtained by by projecting $\hat{\mathbb{W}}_{\mathfrak{F}} (L)$ along the control space onto the state space and is expressed as 
	\begin{eqnarray}
	&\hat{\mathbb{X}}_{\mathfrak{F}} (L) \!=\! \Big\{\mathbb{C}^x \!\!\in\! \{\mathbb{C}^x_i\} \Big| \exists \mathbb{C}^u  \!\!\in\! \{\mathbb{C}^u_j\},\nonumber\\
	& \mathbb{C}^x \!\!\times\! \mathbb{C}^u \!\in\! \hat{\mathbb{W}}_{\mathfrak{F}}(L) \Big\}. \label{eq:hat{x}_F(L)}
	\end{eqnarray}
	From \eqref{eq:hat{W}_F(L)}, it is obvious that $\hat{\mathbb{W}}_{\mathfrak{F}}(L)$ is an inner approximation of $\mathbb{W}_{\mathfrak{F}}(L)$. Since $L(f(0,0)) - L(0) = 0, \forall f \in \mathfrak{F}$, the origin $0 \in \mathbb{R}^{n + m}$ is in boundary of $\mathbb{W}_{\mathfrak{F}}(L)$. Hence, there is no cell belonging to the inner approximation $\hat{\mathbb{W}}_{\mathfrak{F}}(L)$ and there is a small neighborhood of the origin $0 \in \mathbb{R}^n$ which is not contained by the projection $\hat{\mathbb{X}}_{\mathfrak{F}}(L)$ of $\hat{\mathbb{W}}_{\mathfrak{F}}(L)$. The size of the neighborhood is smaller when the size of cells is smaller. However, from \eqref{eq:thm:robust_stab:Omega}-\eqref{eq:thm:robust_stab:X}, we know that $\mathbb{X}_{\mathfrak{F}}(L)$ contains this neighborhood of the origin in the state space except the origin. For convenience of estimating the robust closed-loop DOA, we modify $\hat{\mathbb{X}}_{\mathfrak{F}}(L)$ obtained by \eqref{eq:hat{x}_F(L)} such that it contains this neighborhood of the origin and the origin.
	
	The above procedure of finding $\hat{\mathbb{W}}_{\mathfrak{F}} (L)$ and $\hat{\mathbb{X}}_{\mathfrak{F}}(L)$ is summarized in Algorithm 1.
	
	\begin{algorithm} 
		\caption{Estimation of NDDs for a given Lyapunov function} \label{alg:est_W_F(L)}
		\textbf{Inputs}: 
		
		\ \ -\ positive-definite function $L: \mathbb{R}^n \to \mathbb{R}$;
		
		\ \ -\ plant set $\mathfrak{F}$ defined in \eqref{eq:plant_set}.
		
		\textbf{Outputs}:
		
		\ \ -\ $\hat{\mathbb{W}}_{\mathfrak{F}} (L)$ and $\hat{\mathbb{X}}_{\mathfrak{F}}(L)$.
		
		\textbf{Parameters}:
		
		\ \ -\ interested region $\mathbb{W} \subset \mathbb{R}^{n+m}$;
		
		\ \ -\ partition $\{\mathbb{C}_t^w, t = 1,\cdots,N^C_x\cdot N^C_y\}$ of the region $\mathbb{W}$;
		
		\ \ -\ number $N^d_{\text{xu}}$ of data points in $W^d$;
		
		\ \ -\ number $N^d_{\bar{\text{x}}}$ of data points in $\bar{X}^d_{\mathfrak{F}}(x^d,u^d)$.
		
		\textbf{Steps}: 
		
		\begin{algorithmic}[1]
			
			\State Generate the sample data set $W^d$ of $\mathbb{W}$ by \eqref{eq:W^d}, whose data points are drawn from  the uniform distribution on $\mathbb{W}$.
			
			\State For each data point $(x^d;u^d) \in W^d$, generate the sample data set $\bar{X}^d_{\mathfrak{F}}(x^d,u^d)$ by \eqref{eq:x_f^d}, whose data points are drawn from  the uniform distribution on $\bar{\mathbb{X}}_\mathfrak{F} (x^d,u^d)$.
			
			\State Find the sample data set $W^d_{\mathfrak{F}}(L)$ of $\mathbb{W}_{\mathfrak{F}} (L)$ by \eqref{eq:W^d_F(L)}.
			
			\State Obtain the estimate $\hat{\mathbb{W}}_{\mathfrak{F}} (L)$ of $\mathbb{W}_{\mathfrak{F}} (L)$ by \eqref{eq:hat{W}_F(L)}, combining all cells only containing data points in $W_{\mathfrak{F}}^d(L)$.
			
			\State Obtain the estimate $\hat{\mathbb{X}}_{\mathfrak{F}} (L)$ of $\mathbb{X}_{\mathfrak{F}} (L)$ by \eqref{eq:hat{x}_F(L)}, projecting $\bar{\mathbb{W}}(L) $ along the control space onto the state space.

		\end{algorithmic}
	\end{algorithm}
	
	\subsection{Estimating robust colsed-loop DOA and designing controller}
	
	
	Replacing the robust NDD $\mathbb{X}_{\mathfrak{F}} (L) \subset \mathbb{R}^{n}$ with its estimate $\hat{\mathbb{X}}_{\mathfrak{F}} (L)$, from Theorem~\ref{thm:robust_stab}, we know that any Lyapunov function level-set $\mathbb{X}_{\text{ls}} (L,\alpha)$ satisfying 
	\begin{eqnarray}
	\mathbb{X}_{\text{ls}} (L,\alpha) \subset \hat{\mathbb{X}}_{\mathfrak{F}} (L) \label{eq:X_ls(L,alpha) subset hat{X}_F(L)}
	\end{eqnarray} 
	can be an estimate of closed-loop DOA for all plant in the plant set $\mathfrak{F}$.
	
	Replacing the robust NDD $\mathbb{W}_{\mathfrak{F}} (L) \subset \mathbb{R}^{n+m}$ with its estimate $\hat{\mathbb{W}}_{\mathfrak{F}} (L)$, from Theorem~\ref{thm:robust_stab}, we know that any controller $\mu$ satisfying
	\begin{eqnarray}
	\mu(0) = 0, (x;\mu(x)) \in \hat{\mathbb{W}}_{\mathfrak{F}} (L), \forall x \in \hat{\mathbb{X}}_{\mathfrak{F}} (L)\backslash \{0\}  \label{eq:mu subset hat{W}_F(L)}
	\end{eqnarray}
	can asymptotically stabilize all plant in the plant set $\mathfrak{F}$. A simple way to find a controller $\mu$ satisfying \eqref{eq:mu subset hat{W}_F(L)} is that, first, select a controller training set belonging to $\hat{\mathbb{W}}_{\mathfrak{F}} (L)$; then, obtain the controller $\mu$ with a function estimation method, such as interpolation, neural network, Gaussian processes regression and so on. If the trend of the training data points is smooth enough and $\mu(0) = 0$ is constrained, it can be guaranteed that the controller obtained from the function estimator satisfies \eqref{eq:mu subset hat{W}_F(L)}. 
	
	\section{Enlargement robust closed-loop DOA}
	In Section 3, for a given positive-definite function, an estimate of colosed-loop DOA can be obtained. This section introduce a method of searching for better Lyapunov function to make the estimate of the robust closed-loop DOA as large as possible. 
	
	\subsection{Enlarging robust closed-loop DOA for given Lyapunov function}
	
	From Theorem 1, we know that any $\mathbb{X}_{\mathrm{ls}} (L,\alpha)$ satisfing condition \eqref{eq:thm:robust_stab:X_L_alpha} could be an estimate of the robust closed-loop DOA. For the given $L$, the optimization problem 
	\begin{eqnarray}
	\max_{\alpha \in \mathbb{R}_+}\mathfrak{m}(\mathbb{X}_{\mathrm{ls}} (L,\alpha)) \quad \text{subject to \eqref{eq:thm:robust_stab:X_L_alpha}}   
	\label{eq:maxalpha}
	\end{eqnarray}
	can enlarge the estmate of the robust closed-loop DOA, where $\mathfrak{m}(\mathbb{X}_{\mathrm{ls}} (L,\alpha))$ is the Lebesegue measure of $\mathbb{X}_{\mathrm{ls}} (L,\alpha)$(in Euclidean space, it is the volume of $\mathbb{X}_{\mathrm{ls}} (L,\alpha)$). However, optimization problem \eqref{eq:maxalpha} cannot be solved, because calculating $\mathfrak{m}(\mathbb{X}_{\mathrm{ls}} (L,\alpha))$ and verifying the constraint \eqref{eq:thm:robust_stab:X_L_alpha} are impossible due to nonlinearities of $L$ and $\mathfrak{F}$. In order to handle $\mathfrak{m}(\mathbb{X}_{\mathrm{ls}} (L,\alpha))$ and the set containment constrain \eqref{eq:thm:robust_stab:X_L_alpha}, $\mathbb{X}_{\mathrm{ls}} (L,\alpha)$ and $\mathbb{X}_{\mathfrak{F}} (L)$ are replaced with their estimates $\hat{\mathbb{X}}_{\mathrm{ls}} (L,\alpha)$ and $\hat{\mathbb{X}}_{\mathfrak{F}} (L)$, respectively, and optimization problem \eqref{eq:maxalpha} is rewritten as
	\begin{eqnarray}
	&\max_{\alpha \in \mathbb{R}_+}\mathfrak{m}(\hat{\mathbb{X}}_{\mathrm{ls}} (L,\alpha)) \quad \text{subject to} \label{eq:maxalpha_hat_objective}  \\
	&\qquad \hat{\mathbb{X}}_{\text{ls}} (L,\alpha) \subset \hat{\mathbb{X}}_{\mathfrak{F}} (L), \label{eq:maxalpha_hat_constrain}
	\end{eqnarray}
	where $\hat{\mathbb{X}}_{\mathfrak{F}} (L)$ is obtained by Algorithm~\ref{alg:est_W_F(L)} and $\hat{\mathbb{X}}_{\mathrm{ls}} (L,\alpha)$ can be obtained with the same idea of estimating $\mathbb{W}_{\mathfrak{F}} (L)$ in Algorithm~\ref{alg:est_W_F(L)}. First, a sample data set $X^d$ of the interested region $\mathbb{X} \subset \mathbb{R}^n$ in the state space is generated, in which each data point $x^d \in \mathbb{R}^n$ is drawn from the uniform distribution on $\mathbb{X}$ and the number of data points is $N^d_x$. Then, the sample data set 
	\begin{eqnarray}
	X^d_{\mathrm{ls}}(L,\alpha) := \Big\{x^d \in X^d \Big| L(x^d) \leq \alpha \Big\}
	\end{eqnarray}
	of $\mathbb{X}_{\mathrm{ls}} (L,\alpha)$ can be obtained. Finally, with the partition $\{\mathbb{C}^x_i, i = 1,2,\cdots,N^C_x\}$ of $\mathbb{X}$ which is same as the one in estimation of $\mathbb{X}_{\mathfrak{F}} (L)$, the estimate $\hat{\mathbb{X}}_{\mathrm{ls}} (L,\alpha)$ of $\mathbb{X}_{\mathrm{ls}} (L,\alpha)$ can be obtained by combining all cells only containing data points in $X^d_{\mathrm{ls}}(L,\alpha)$, namely, 
	\begin{eqnarray}
	&\hat{\mathbb{X}}_{\mathrm{ls}} (L,\alpha) = \Big\{\mathbb{C}^x \in \{\mathbb{C}^x_i\} \Big| \forall x^d \in \mathbb{C}^x \cap X^d, \nonumber\\ 
	&x^d \in X^d_{\mathrm{ls}}(L,\alpha) \Big\}. \label{eq:est_xls}
	\end{eqnarray}
	Due to that $\hat{\mathbb{X}}_{\mathrm{ls}} (L,\alpha)$ consists of cells, it is easily to calculate the volume of $\hat{\mathbb{X}}_{\mathrm{ls}} (L,\alpha)$, namely, $\mathfrak{m}(\hat{\mathbb{X}}_{\mathrm{ls}} (L,\alpha))$ in \eqref{eq:maxalpha_hat_objective}. Due to that both $\hat{\mathbb{X}}_{\mathrm{ls}} (L,\alpha)$ and $\hat{\mathbb{X}}_{\mathfrak{F}} (L)$ consist of cells, the set containment constrain \eqref{eq:maxalpha_hat_constrain} is equivalent to 
	\begin{eqnarray}
	X^C_{\text{ls}} (L,\alpha) \cap X^C_{\mathfrak{F}} (L)=X^C_{\text{ls}} (L,\alpha) \label{eq:cap}
	\end{eqnarray}
	where $X^C_{\text{ls}} (L,\alpha)\in\{0,1\}^{N^C_x}$ and $X^C_{\mathfrak{F}}\in\{0,1\}^{N^C_x}$ are logical vectors and the binary operator $\cap$ represents the logical operator AND for two logical vectors element by element. Each cell in $\{\mathbb{C}^x_i\}$ corresponds to a element of $X^C_{\text{ls}} (L,\alpha)$ and $X^C_{\mathfrak{F}} (L)$. If a cell belongs to $\hat{\mathbb{X}}_{\mathrm{ls}} (L,\alpha)$, the corresponding element of $X^C_{\text{ls}} (L,\alpha)$ equals to 1, otherwise equals to 0. If a cell belongs to $\hat{\mathbb{X}}_{\mathfrak{F}} (L)$, the corresponding element of $X^C_{\mathfrak{F}}(L)$ equals to 1, otherwise equals to 0. 
	
	When $\mathfrak{m}(\hat{\mathbb{X}}_{\mathrm{ls}} (L,\alpha))$ can be calculated and the set containment constrain \eqref{eq:maxalpha_hat_constrain} can be verified, optimization problem \eqref{eq:maxalpha_hat_objective}-\eqref{eq:maxalpha_hat_constrain} can be solved. Note that the volume of $\hat{\mathbb{X}}_{\mathrm{ls}} (L,\alpha)$ is increasing as $\alpha$ is increasing for the given $L$. The idea of solving optimization problem \eqref{eq:maxalpha_hat_objective}-\eqref{eq:maxalpha_hat_constrain} is very simple. Initially set $\alpha := \epsilon$, and repeat $\alpha := \alpha+\epsilon$ until the set containment constrain \eqref{eq:maxalpha_hat_constrain} is not satisfied, where $\epsilon \in \mathbb{R}_+$ is a given small constant. Let $\alpha^\ast(L)$ denotes the solution of optimization problem \eqref{eq:maxalpha_hat_objective}-\eqref{eq:maxalpha_hat_constrain}. The above procedure finding $\alpha^\ast(L)$ is summarized in Algorithm~\ref{alg:alpha}. In order to improve the search efficiency, Algorithm~\ref{alg:alpha} uses the variable $\epsilon$ scheme rather than the constant $\epsilon$ scheme. For example, suppose $\alpha^\ast(L) = 97.678$, for the constant $\epsilon = 0.001$ scheme, the number of verifying constrain \eqref{eq:maxalpha_hat_constrain} is 97679; for the variable $\epsilon$ scheme, where $\epsilon^{init} = 10$ and the desired accuracy is $0.001$, the number of verifying constrain \eqref{eq:maxalpha_hat_constrain} is 42. Generally, set $\epsilon^{init} = 10^\eta, \eta \in \mathbb{Z}_+$. After obtaining $\alpha^\ast(L)$, the largest estimate of the robust closed-loop DOA for the given Lyapunov function $L$ is $\hat{\mathbb{X}}_{\mathrm{ls}} (L,\alpha^\ast(L))$.
	
	\begin{algorithm} 
		\caption{Finding the optimal $\alpha^\ast(L)$ for the given Lyapunov function $L$} \label{alg:alpha}
		\textbf{Inputs}: 
		
		\ \ -\ positive-definite function $L: \mathbb{R}^n \to \mathbb{R}$;
		
		\ \ -\ estimate $\hat{\mathbb{X}}_{\mathfrak{F}} (L)$ of NDD $\mathbb{X}_{\mathfrak{F}} (L)$.
		
		\textbf{Outputs}:
		
		\ \ -\ solution $\alpha^\ast(L)$ of optimization problem \eqref{eq:maxalpha_hat_objective}-\eqref{eq:maxalpha_hat_constrain};
		
		\ \ -\ largest $\hat{\mathbb{X}}_{\mathrm{ls}} (L,\alpha^\ast(L))$ satisfying \eqref{eq:maxalpha_hat_constrain} for given $L$.
		
		\textbf{Parameters}:
		
		\ \ -\ $\epsilon^{init}$
		
		\ \ -\ desired accuracy
		
		\textbf{Steps}: 
		
		\begin{algorithmic}[1]
			\State Set $\epsilon := \epsilon^{init}$ and $\alpha := \epsilon$, where $\epsilon^{init} \in \mathbb{R}_+$ is a constant.
			
			\State Find estimate $\hat{\mathbb{X}}_{\mathrm{ls}} (L,\alpha)$ of $\mathbb{X}_{\mathrm{ls}} (L,\alpha)$ by \eqref{eq:est_xls}.
			
			\State Verify constrain \eqref{eq:maxalpha_hat_constrain} by \eqref{eq:cap}, if constrain \eqref{eq:maxalpha_hat_constrain} is satisfied, then set $\alpha:=\alpha+\epsilon$ and go to Step2.
			
			\State Set $\alpha:=\alpha-\epsilon$ and $\epsilon:=0.1\cdot\epsilon$.
			
			\State If $\epsilon$ satisfies the desired accuracy, then set $\alpha^\ast(L) := \alpha$ and $\hat{\mathbb{X}}_{\mathrm{ls}} (L,\alpha^\ast(L)) := \hat{\mathbb{X}}_{\mathrm{ls}} (L,\alpha)$, otherwise $\alpha:=\alpha+\epsilon$ and go to Step2.
			
		\end{algorithmic}
	\end{algorithm}
	
	\subsection{Enlarging robust closed-loop DOA by selecting appropr \newline -iate Lyapunov function}
	
	When the Lyapunov function $L: \mathbb{R}^n \to \mathbb{R}$ is not given but can be selected from a positive-definite function set, a significantly larger estimate of the robust closed-loop DOA may be obtained by solving the following optimization problem.
	\begin{eqnarray}
	\max_{L\in\mathfrak{C}_n}\mathfrak{m}(\hat{\mathbb{X}}_{\mathrm{ls}} (L,\alpha^\ast(L))),  \label{eq:maxL}
	\end{eqnarray}  
	where $\mathfrak{C}_n$ denotes the set of all continuous positive-definite functions in $n$ variables and $\hat{\mathbb{X}}_{\mathrm{ls}} (L,\alpha^\ast(L))$ is the largest estimate of the robust closed-loop DOA for $L$. Unfortunately, optimization problem \eqref{eq:maxL} is an infinite dimensional problem and unsolvable. Hence, we restrict $L\in\mathfrak{L}_{n,2d}$ rather than $L\in\mathfrak{C}_n$. Function set $\mathfrak{L}_{n,2d}$ is a subset of all sum-of-square polynomials and defined as
	\begin{eqnarray}
	\mathfrak{L}_{n,2d}=\Big\{L\in\mathfrak{R}_{n,2d}\Big|L(x)=S^T_d(x)Q^TQS_d(x)  \label{eq:generate L}\Big\}
	\end{eqnarray}  
	where $\mathfrak{R}_{n,2d}$ is the set of all polynomials in $n$ variables with degree $\leq2d$, $x = (x_{(1)};x_{(2)};\cdots;x_{(n)}) \in \mathbb{R}^n$, $Q\in\mathbb{R}^{r\times r}$ is a full rank matrix, $S_d(x)=(x_{(1)};...;x_{n};x_{(1)}x_{(2)};...;x^d_{(n)}) \newline \in \mathbb{R}^r$ and $r = (_{{\kern 1pt} {\kern 1pt} {\kern 1pt} {\kern 1pt} d}^{n + d}) - 1$. $\mathfrak{L}_{n,2d}$ is choosen because, $\forall L\in\mathfrak{L}_{n,2d}$, $L$ is positive-definite, namely $L(0)=0$ and $L(x)>0$, $\forall x\in\mathbb{R}^n\backslash\{0\}$. With $\mathfrak{L}_{n,2d}$, the unsolvable optimization problem \eqref{eq:maxL} is rewritten as the following solvable optimization problem.
	\begin{eqnarray}
	\max_{L\in\mathfrak{L}_{n,2d}}\mathfrak{m}(\hat{\mathbb{X}}_{\mathrm{ls}} (L,\alpha^\ast(L)))
	\label{eq:max_est_L}
	\end{eqnarray}  
	
	The positive-definite function set $\mathfrak{L}_{n,2d}$ is a parameterized sum-of-square polynomial function set, whose parameters is $Q\in\mathbb{R}^{r\times r}$. We define function $m: \mathbb{R}^{r \times r} \to \mathbb{R}_+$ as 
	$
	m(Q)=\mathfrak{m}(\hat{\mathbb{X}}_{\mathrm{ls}} (L,\alpha^{\ast}(L))), \nonumber
	$
	where $L(x)=S^T_d(x)Q^TQS_d(x)$ and $\alpha^{\ast}(L)$ obtained by Algorithm~\ref{alg:alpha}. With the function $m(Q)$ ,the optimization problem \eqref{eq:max_est_L} can be equivalently rewritten as
	\begin{eqnarray}
	\max_{Q\in\mathbb{R}^{r\times r}}m(Q).  \label{eq:max Q}
	\end{eqnarray}  
	The analytical expression of $m(Q)$ is hard to be derived, but it is easy to evaluate $m(Q)$ for the given $Q$. The procedure of evaluating $m(Q)$ for the given $Q$ is summarized in Algorithm~\ref{alg:m(Q)}. Hence, classic optimization methods, \textit{e.g.}, gradient descent method, cannot be used to solve the optimization problem \eqref{eq:max Q}. However, meta-heuristic optimization methods can be used to solve the optimization problem \eqref{eq:max Q}, whose advantage is that the function to be optimized is only required to be evaluable. Popular meta-heuristic optimizers for real-valued search-spaces include particle swarm optimization, differential evolution and evolution strategies. There are lots of literatures about meta-heuristic optimizers, so we omit an introduction about them in this paper.
	
	\begin{algorithm} 
		\caption{evaluating $m(Q)$ for given $Q$} \label{alg:m(Q)}
		\textbf{Inputs}: 
		
		\ \ -\ $Q \in \mathbb{R}^{r \times r}$
		
		\ \ -\ plant set $\mathfrak{F}$ defined in \eqref{eq:plant_set};

		\textbf{Outputs}:
		
		\ \ -\ $m(Q) \in \mathbb{R}_+$;

		\textbf{Steps}: 
		
		\begin{algorithmic}[1]
			\State Determine the positive-definite function $L(x) = S^T_d(x)Q^TQS_d(x)$ according to $Q$.
			
			\State Based on $L$, $\hat{f}$, $\delta$ and $\{\mathbb{C}^w_t\}$, obtain $\hat{\mathbb{X}}_{\mathfrak{F}}(L)$ by Algorithm~\ref{alg:est_W_F(L)}.
			
			\State Based on $L$  and $\mathfrak{F}$, obtain $\hat{\mathbb{X}}_{\mathrm{ls}} (L,\alpha^{\ast}(L))$ by Algorithm~\ref{alg:alpha}.
			
			\State Obtain ${m}(Q)=\mathfrak{m}(\hat{\mathbb{X}}_{\mathrm{ls}} (L,\alpha^{\ast}(L)))$.

		\end{algorithmic}

	\end{algorithm}
	
	\section{Simulation}
	
	Consider the nominal model $\hat{f}$ and the modeling error bound $\delta(x,u)$:
	\begin{eqnarray}
	\hat{f}(x,u) &=& -\sin(2x) - xu - 0.2x - u^2 + u, \nonumber \\
	\delta(x,u) &=& 1 - \exp \left(-0.5(x^2 + u^2)\right), \nonumber
	\end{eqnarray} 
	where $x \in \mathbb{R}$ and $u \in \mathbb{R}$. The interested region $\mathbb{W}=[-2,2]\times[-2,2]\subset\mathbb{R}^2$ in the state-control space is partitioned into $1.6\times10^5$ cells of size $0.01\times0.01$. The number of data points in $W^d$ is selected as $5\times 10^6$. For each data point $(x^d;u^d)$ in $W^d$, the number of data points in  $\bar{\mathbb{X}}_\mathfrak{F} (x^d,u^d)$ is selected as 500.
	
\begin{figure*}
		\begin{center}
			\includegraphics[width=0.29\textwidth]{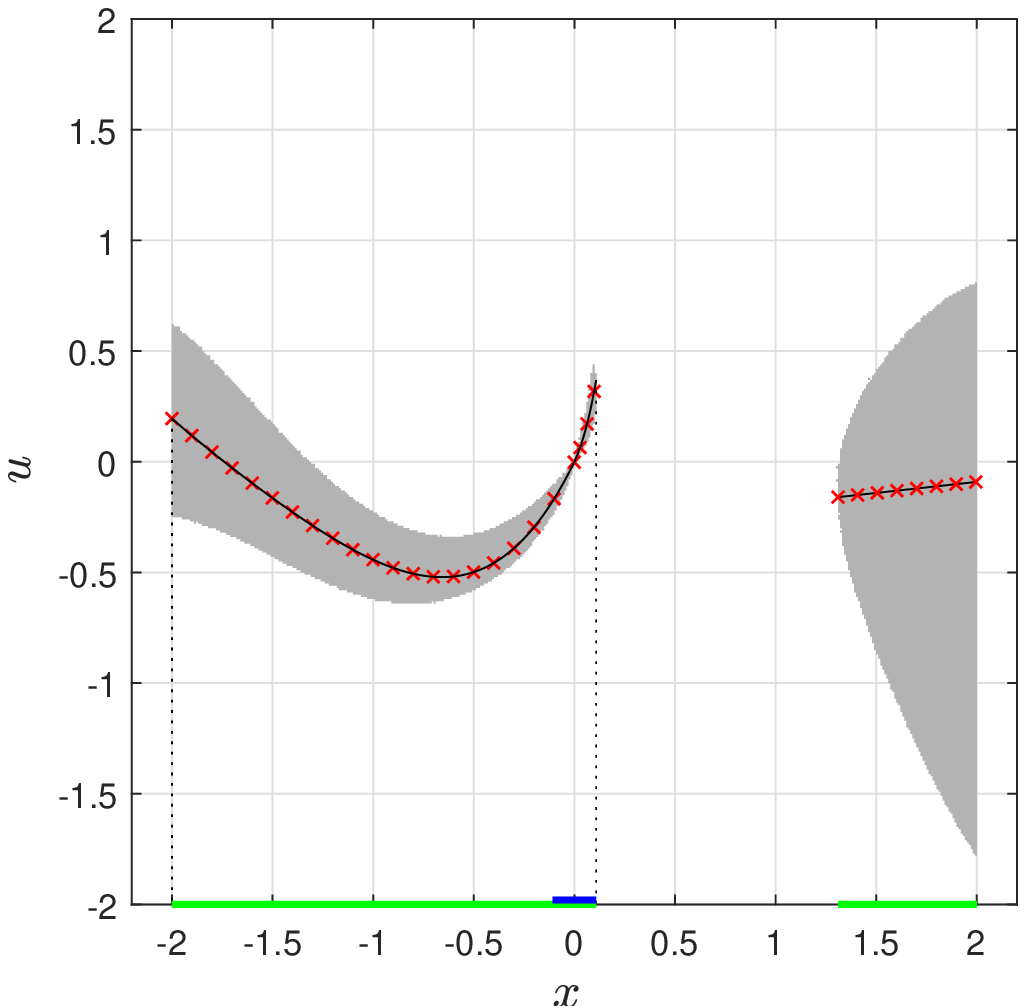}
			\includegraphics[width=0.29\textwidth]{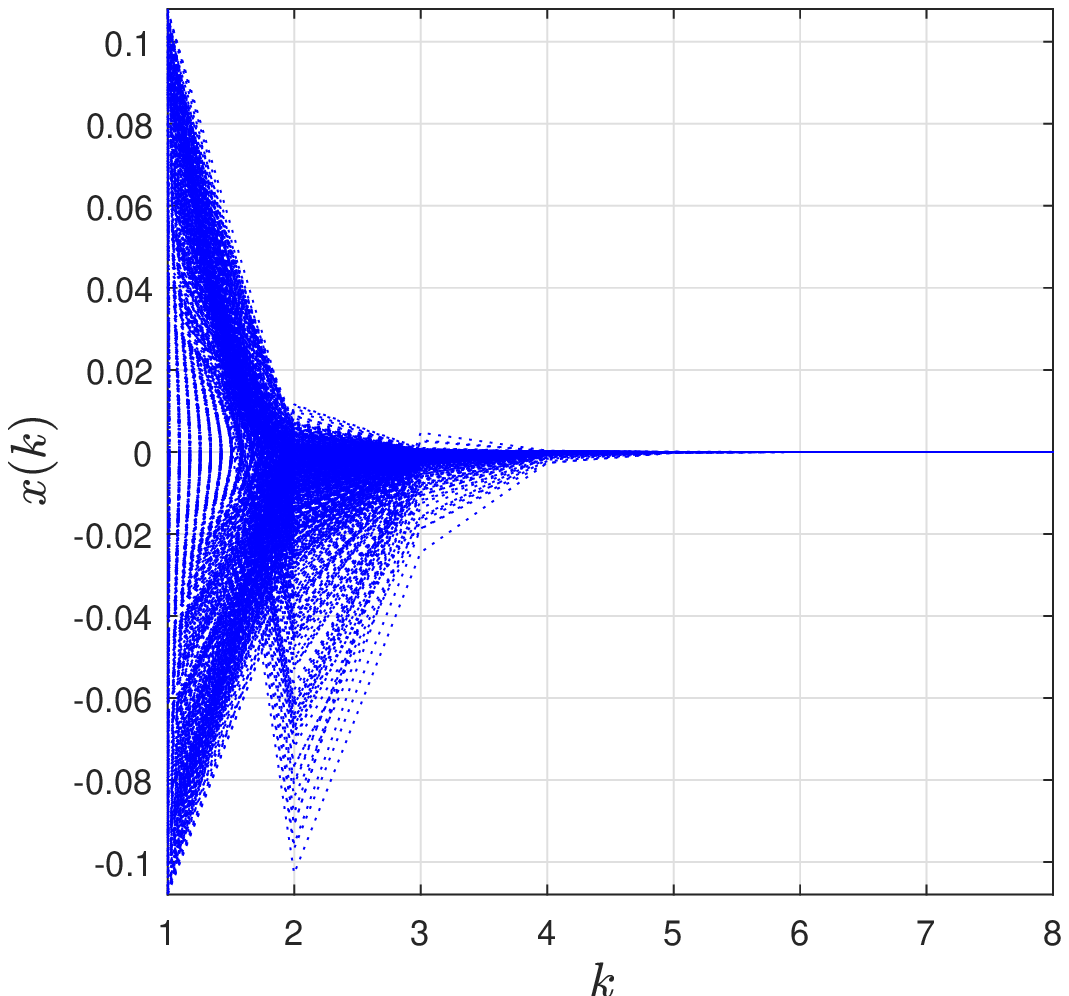}
			\includegraphics[width=0.29\textwidth]{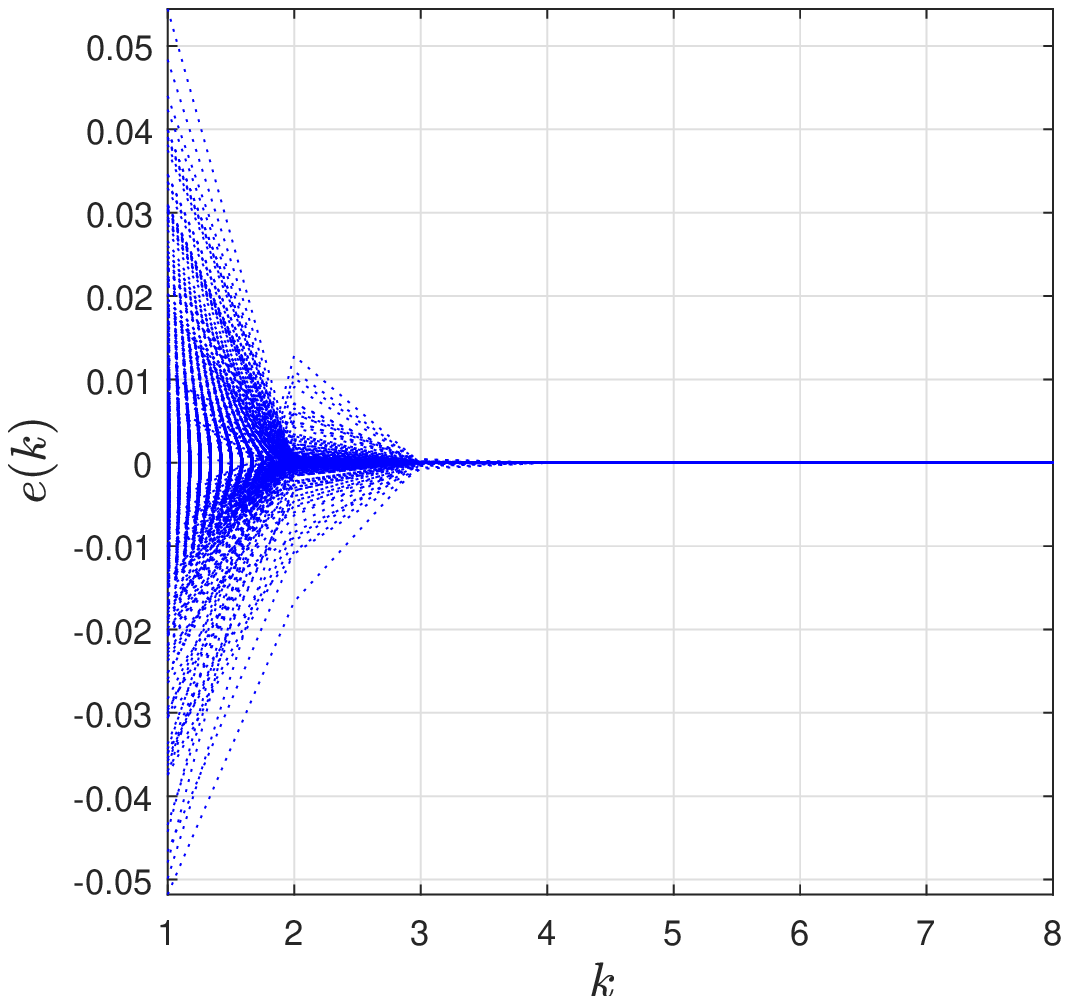}\\
			\parbox[c]{0.29\textwidth}{\footnotesize \centering (a)}
			\parbox[c]{0.29\textwidth}{\footnotesize \centering (b)}
			\parbox[c]{0.29\textwidth}{\footnotesize \centering (c)}
			\caption{(a) Estimates $\hat{\mathbb{W}}_{\mathfrak{F}}(L),\hat{\mathbb{X}}_\mathfrak{F}(L)$ of robust NDDs, estimate $\hat{\mathbb{X}}_{\mathrm{ls}}(L,0.0117)$ of closed-loops, controller training data and robust controller $\mu$. (b) State trajectories of closed-loops $x(k)$. (c) Noise trajectories $e(k)$.}
			\label{fig:controller}
		\end{center}
	\end{figure*}

	When the positive-definite function $L(x) = x^2$ is selected, an estimate $\hat{\mathbb{W}}_\mathfrak{F}(L)$ of the robust NDD $\mathbb{W}_\mathfrak{F}(L)$ is obtained by Algorithm~\ref{alg:est_W_F(L)} and shown in Fig~\ref{fig:controller}(a) denoted by gray region. An estimate $\hat{\mathbb{X}}_\mathfrak{F}(L)$ of the robust NDD $\mathbb{X}_\mathfrak{F}(L)$ is also obtained and shown in Fig~\ref{fig:controller}(a) denoted by green line segment in x-axis. The largest estimate 
	\begin{eqnarray}
	\hat{\mathbb{X}}_{\mathrm{ls}} (x^2,0.0117)=[-0.108,0.108]\subset\mathbb{R} \nonumber
	\end{eqnarray} 
	of the robust closed-loop DOA for $L(x) = x^2$ is found by Algorithm~\ref{alg:alpha} and shown in Figure~\ref{fig:controller}(a) denoted by the blue line segment in x-axis. In order to find a controller $\mu$ belonging to the gray region, we select a training data set shown by red $'\times'$ in Figure~\ref{fig:controller} (a). With the training data set, the robust controller $\mu$ is obtained using Gaussian processes regression, as shown in Figure~\ref{fig:controller} denoted by black line. To verify whether the controller $\mu$ can stabilize all plants in the plant set for all initial state in $\hat{\mathbb{X}}_{\mathrm{ls}} (x^2,0.0117)$, we consider the controlled system $x(k+1)=\hat{f}(x(k),u(k))+e(k)$, where noise $e(k)$ is drawn from the uniform distribution on $[-\delta(k),\delta(k)]\subset\mathbb{R}$ and $\delta(k)=\delta(x(k),u(k))$. Figure~\ref{fig:controller}(b) shows 1000 state trajectories of $x(k+1)=\hat{f}(x(k),\mu(x(k)))+e(k)$ denoted by blue dash lines, whose initial states are drawn from the uniform distribution on $\hat{\mathbb{X}}_{\mathrm{ls}} (x^2,0.0117)$. Figure \ref{fig:controller}(c) also shows 1000 noises trajectories corresponding to the 1000 state trajectories. We see that all state trajectories converge to the origin. 
	
	\begin{figure*}
		\begin{center}
			\includegraphics[width=0.30\textwidth]{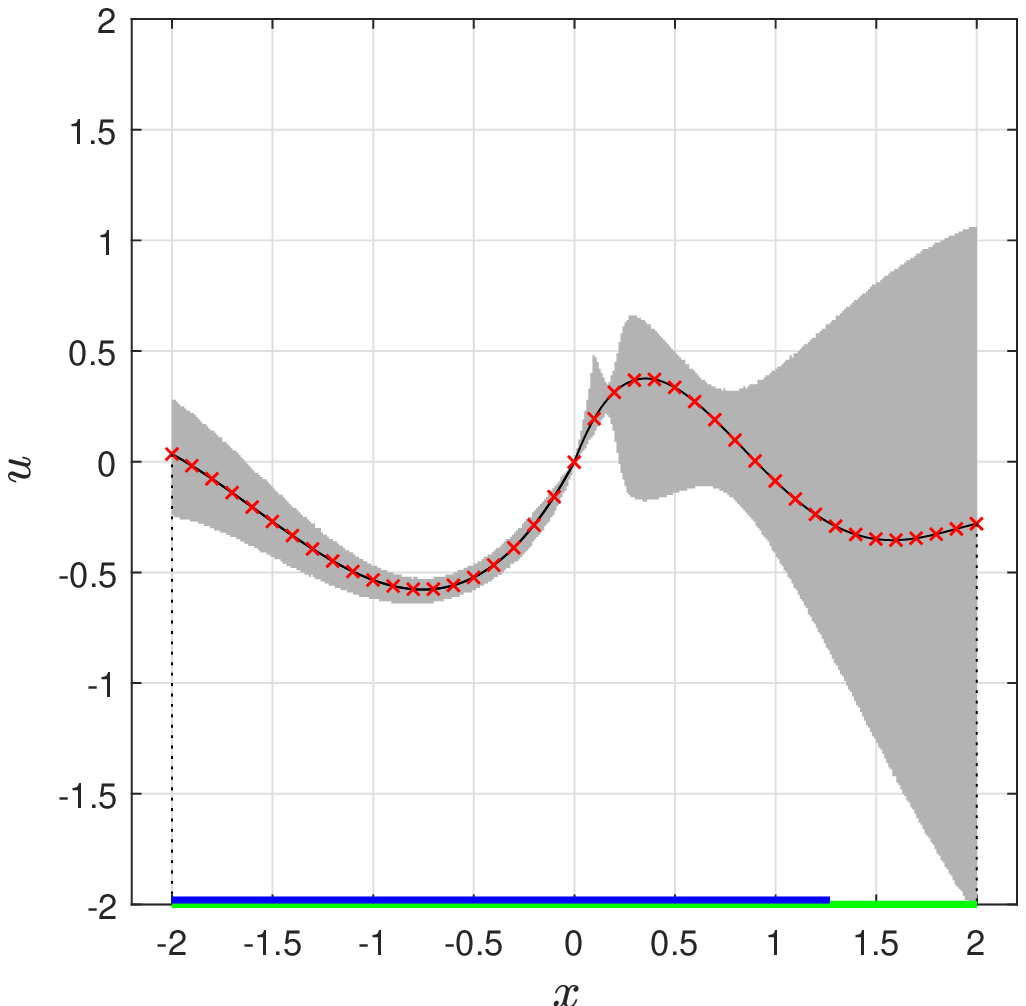}
			\includegraphics[width=0.30\textwidth]{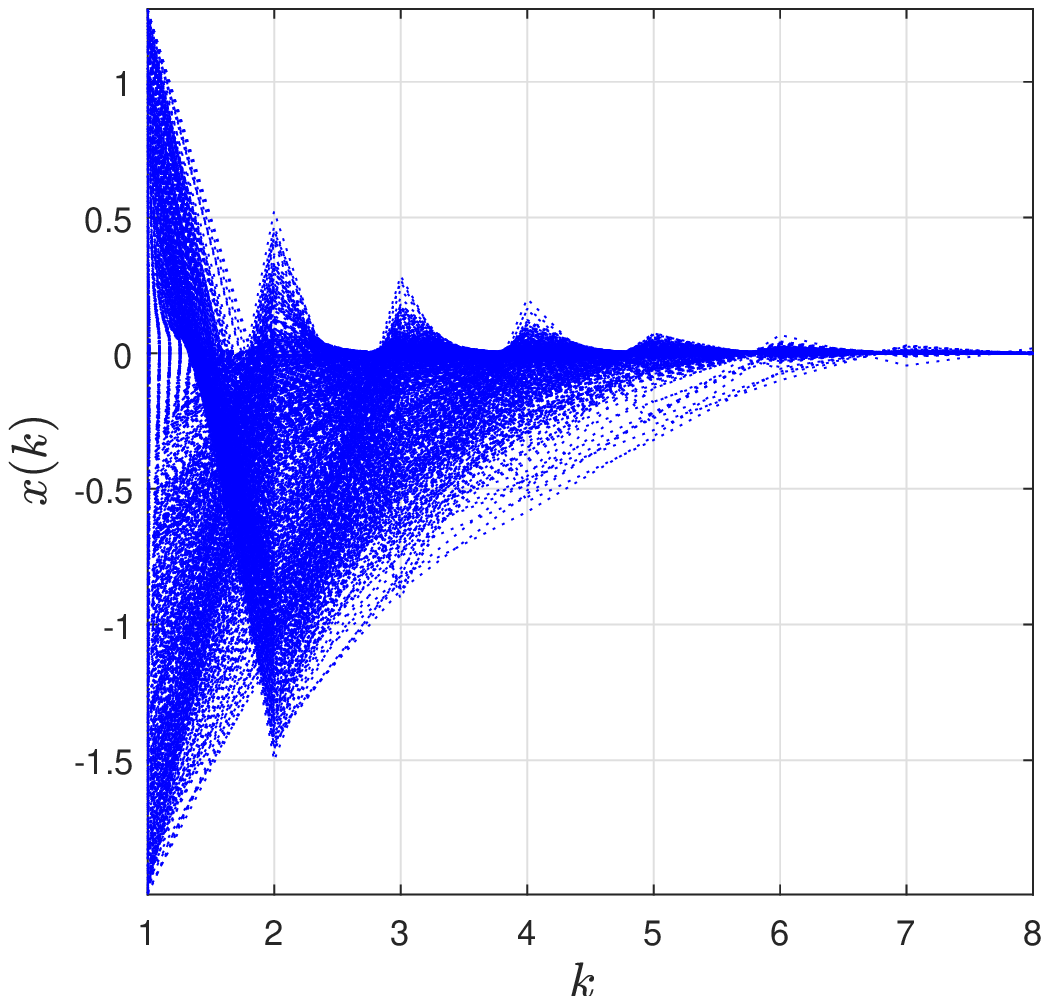}
			\includegraphics[width=0.30\textwidth]{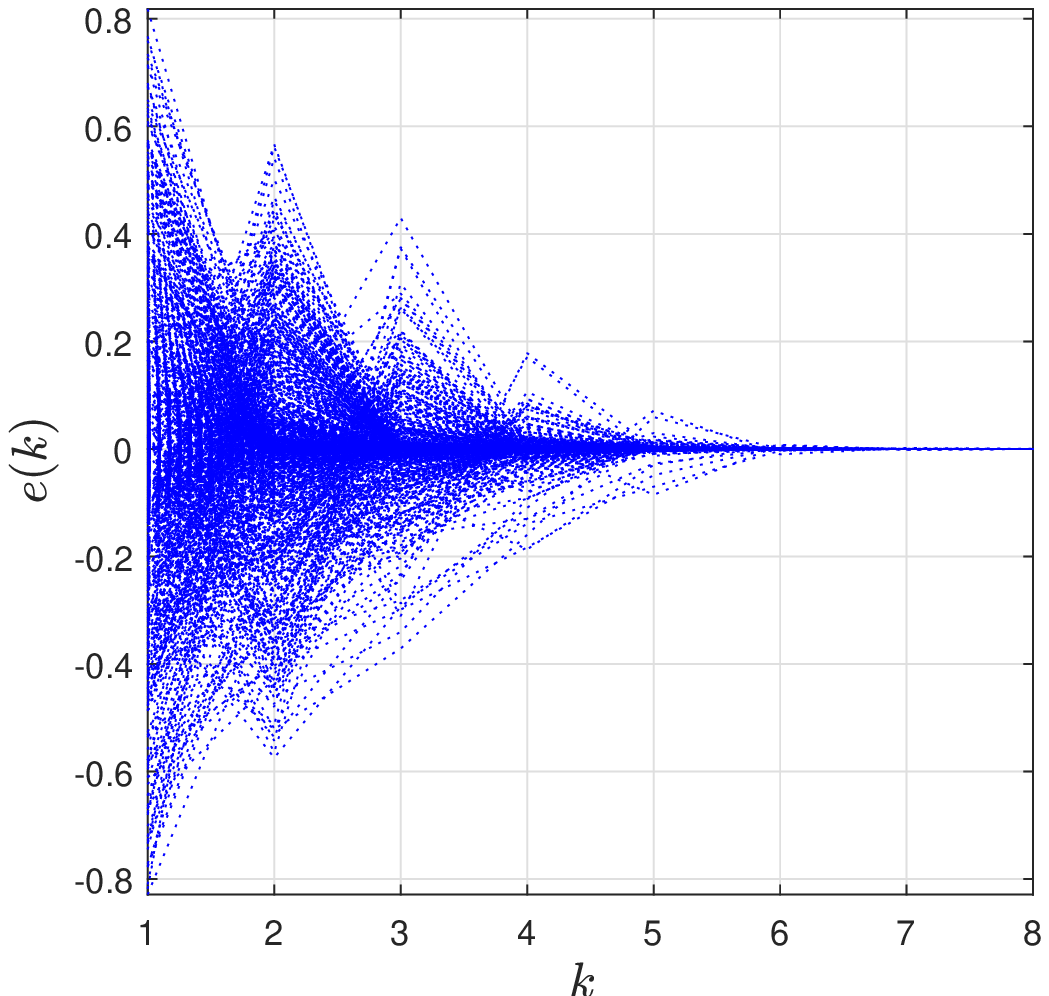}\\
			\parbox[c]{0.30\textwidth}{\footnotesize \centering (a)}
			\parbox[c]{0.30\textwidth}{\footnotesize \centering (b)}
			\parbox[c]{0.30\textwidth}{\footnotesize \centering (c)}
			\caption{(a) Estimates $\hat{\mathbb{W}}_{\mathfrak{F}}(L),\hat{\mathbb{X}}_\mathfrak{F}(L)$ of robust NDDs, estimate $\hat{\mathbb{X}}_{\mathrm{ls}}(L,10.5421)$ of closed-loops, controller training data and robust controller $\mu$. (b) State trajectories of closed-loops $x(k)$. (c) Noise trajectories $e(k)$.}
			\label{fig:doa}
		\end{center}
	\end{figure*}
	
	When the positive-definite function set
	\begin{eqnarray}
	\mathfrak{L}_{1,4}=\Big\{L\in\mathfrak{R}_{1,4}\Big|L(x) = (x;x^2)^TQ^TQ(x;x^2)  \Big\} \nonumber
	\end{eqnarray}
	is selected, the optimization problem \eqref{eq:max Q} is solved through the particle swarm optimization method, where $Q \in \mathbb{R}^{2\times2}$. The solution $Q^\ast = (_{{\rm{1}}{\rm{.0000}}}^{{\rm{0}}{\rm{.3587}}}{\kern 1pt} {\kern 1pt} {\kern 1pt} _{{\rm{0}}{\rm{.8249}}}^{{\rm{0}}{\rm{.9232}}})$. According to $Q^\ast$, the appropriate Lyapunov function $L^\ast(x)=1.5327{x^4} + 2.3121{x^3} + 1.1286{x^2}$ and $\alpha^\ast(L^\ast) = 10.5421$. The estimate $\hat{\mathbb{W}}_\mathfrak{F}(L^\ast)$ of the robust NDD $\mathbb{W}_\mathfrak{F}(L^\ast)$ is shown in Fig~\ref{fig:doa} (a) denoted by gray region. The estimate $\hat{\mathbb{X}}_\mathfrak{F}(L^\ast) = [-2,2] \subset \mathbb{R}$ of the robust NDD $\mathbb{X}_\mathfrak{F}(L^\ast)$ is also shown in Fig~\ref{fig:doa}(a) denoted by green line segment in x-axis. The largest estimate \begin{eqnarray}
	\hat{\mathbb{X}}_{\mathrm{ls}} (L^\ast,\alpha^\ast(L^\ast))=[-2,1.2699]\subset\mathbb{R} \nonumber
	\end{eqnarray}
	of the robust closed-loop DOA for $L^\ast$ is shown in Figure~\ref{fig:doa}(a) denoted by the blue line segment in x-axis. In order to find a controller $\mu$ belonging to the gray region, training data set shown by red $'\times'$ in Figure~\ref{fig:doa}(a) is selected. With the training data set, the robust controller $\mu$ is obtained using Gaussian processes regression, as shown in Figure~\ref{fig:doa}(a) denoted by black line. To verify whether the controller $\mu$ can stabilize all plants in the plant set for all initial state in $\hat{\mathbb{X}}_{\mathrm{ls}} (L^\ast,\alpha^\ast(L^\ast))$, the same control system $x(k+1)=\hat{f}(x(k),u(k))+e(k)$ is considered. Figuer~\ref{fig:doa}(b) shows 1000 state trajectories of $x(k+1)=\hat{f}(x(k),\mu(x(k)))+e(k)$ denoted by blue dash lines, whose initial states are drawn from the uniform distribution on $\hat{\mathbb{X}}_{\mathrm{ls}} (L^\ast,\alpha^\ast(L^\ast))$.  Figure~\ref{fig:doa}(c) also shows 1000 noises trajectories corresponding to the 1000 state trajectories. It shows that all state trajectories converge to the origin. As shown in Fig~\ref{fig:doa}(a), the new DOA is $[-2.0,1.2699]$, compared with Fig~\ref{fig:controller}, the DOA has enlarged by 15.1384 times.

	\section{Conclusion}
	This paper proposed a new data-driven control method to asymptotic stabilize non-affine nonlinear plant with DOA enlargement. By applying Lyapunov approach, a solvable optimization problem was formulated to enlarge robust closed-loop DOA iterated through parameterized positive-definite function set. Simulation results verified the effectiveness of  all algorithms in the method. The sufficient conditions are stringent, finding looser conditions is our future work.

	\bibliography{robust_doa_enlarge}             
	
	
	
	
	
	
	
	
	\appendix

\end{document}